\documentclass[aps,footinbib,prd,superscriptaddress,reprint]{revtex4-1}

\usepackage{graphicx}% Include figure files
\usepackage{amssymb}

\usepackage{amsmath}
\usepackage{slashed}
\usepackage{lipsum}
\usepackage[dvipsnames]{xcolor}
\usepackage{xcolor}   
\usepackage{xspace} 
\usepackage{bm}
\usepackage{xfrac}
\usepackage{lineno}
\usepackage[normalem]{ulem}

\usepackage[T1]{fontenc}      
\usepackage{ae} 
\usepackage{grffile}

\newcommand{\im}{\ensuremath{\mathrm{i}}}

\usepackage[colorlinks=true]{hyperref}
\hypersetup{
    bookmarks=true,         % show bookmarks bar?
    unicode=false,          % non-Latin characters
    pdftoolbar=true,        % show Acrobat
    pdfmenubar=true,        % show Acrobat
    pdffitwindow=false,     % window fit to page when opened
    pdfstartview={FitH},    % fits the width of the page to the window
    pdfnewwindow=true,      % links in new window
    colorlinks=true,        % false: boxed links; true: colored links
    linkcolor=blue,         % color of internal links (change box color with linkbordercolor)
    citecolor=red,          % color of links to bibliography
    filecolor=magenta,      % color of file links
    urlcolor=blue           % color of external links
}

\begin{document}

\title{Temperature and Chemical Potential Dependence of the Parity Anomaly in Quantum Anomalous Hall Insulators}

%\title{Temperature and Chemical Potential Dependence of the Parity Anomaly in Quantum Anomalous Hall Insulators}

%\author[a,d]{C.~Tutschku}
%\author[b,d]{F.~S.~Nogueira}
%\author[a,d]{C.~Northe}
%\author[b,c,d]{J.~van~den~Brink}
%\author[a,d]{E.~M.~Hankiewicz} 

\affiliation{Institute for Theoretical Physics,
Julius-Maximilians-Universit\"at W\"urzburg, 97074 W\"urzburg, Germany}

\affiliation{Institute for Theoretical Solid State Physics, IFW Dresden, 01069 Dresden, Germany}

\affiliation{Institute for Theoretical Physics, TU Dresden, 01069 Dresden, Germany}

\affiliation{W\"urzburg-Dresden Cluster of Excellence ct.qmat}

\author{Christian~Tutschku}

\affiliation{Institute for Theoretical Physics,
Julius-Maximilians-Universit\"at W\"urzburg, 97074 W\"urzburg, Germany}

\affiliation{W\"urzburg-Dresden Cluster of Excellence ct.qmat}

\author{Flavio~S.~Nogueira}

\affiliation{Institute for Theoretical Solid State Physics, IFW Dresden, 01069 Dresden, Germany}

\affiliation{W\"urzburg-Dresden Cluster of Excellence ct.qmat}

\author{Jeroen~van~den~Brink}

\affiliation{Institute for Theoretical Solid State Physics, IFW Dresden, 01069 Dresden, Germany}

\affiliation{Institute for Theoretical Physics, TU Dresden, 01069 Dresden, Germany}

\affiliation{W\"urzburg-Dresden Cluster of Excellence ct.qmat}

\author{E.~M.~Hankiewicz} 

\affiliation{Institute for Theoretical Physics,
Julius-Maximilians-Universit\"at W\"urzburg, 97074 W\"urzburg, Germany}

\affiliation{W\"urzburg-Dresden Cluster of Excellence ct.qmat}

\begin{abstract}
The low-energy physics of two-dimensional Quantum Anomalous Hall insulators like (Hg,Mn)Te quantum wells or magnetically doped (Bi,Sb)Te thin films can be effectively described by  two Chern insulators, including a Dirac, as well as a momentum-dependent mass term. Each of those Chern insulators is directly related to the parity anomaly of planar quantum electrodynamics. In this work, we analyze the finite temperature Hall conductivity of a single  Chern insulator in 2+1 space-time dimensions under the  influence of a chemical potential and an out-of-plane magnetic field. At zero magnetic field, this non-dissipative transport coefficient originates from the parity anomaly of planar quantum electrodynamics. We show that the parity anomaly itself is not renormalized by finite temperature effects. However, it induces two terms of different physical origin in the effective action of a Chern insulator, which is proportional to the Hall conductivity. The first term is temperature and chemical potential independent, and solely encodes the intrinsic topological response. The second term specifies the non-topological thermal response of conduction and valence band states. In particular, we show that the relativistic mass of a Chern insulator counteracts finite temperature effects, whereas its non-relativistic mass enhances these corrections.  Moreover, we  extend our analysis to finite magnetic fields and relate the thermal response of a Chern insulator therein to the spectral asymmetry, which is a measure of the parity anomaly in orbital fields.
\end{abstract}

%\emailAdd{christian.tutschku@physik.uni-wuerzburg.de}
%\emailAdd{f.de.souza.nogueira@ifw-dresden.de}
%\emailAdd{christian.northe@physik.uni-wuerzburg.de}
%\emailAdd{j.van.den.brink@ifw-dresden.de}
%\emailAdd{Ewelina.Hankiewicz@physik.uni-wuerzburg.de}

%\keywords{Parity Anomaly, Quantum Anomalous Hall Effect, Spectral Asymmetry,\\ Finite Temperature and Density Quantum Field Theory,  Topological States of Matter}

%\arxivnumber{To be added}

\maketitle

%\flushbottom 

\section{Introduction} \label{Sec1}

%\vspace{-.3cm}

Back in the 1980s, Haldane proposed the first solid state model of a Quantum Anomalous Hall (QAH) insulator by adding a parity-breaking \footnote{Instead of a parity-even Dirac mass term, initially studied by Semenoff in Ref.~\cite{Semenoff84}.} Dirac mass term to an otherwise gapless graphene structure \cite{Haldane88}. Such a system features a non-zero Hall conductivity even in the absence of Landau levels (LLs). From a high-energy perspective, this model 
is directly related to the parity-anomaly of planar quantum electrodynamics, which implies that it is not possible to quantize a single Dirac fermion in a parity symmetric manner \cite{Niemi83,Redlich83}. Strictly speaking, the Haldane model contains two Dirac fermions as it is based on the hexagonal lattice structure of graphene. However, by fine-tuning the Haldane mass,  one of the Dirac fermion mass gaps can be closed, whereas the other one remains open. In this parameter limit, the  band-structure contains a single gapless Dirac fermion with a non-zero Hall conductivity. This implies that in the Haldane model one of the Dirac fermions alone is suitable to  realize the parity anomaly in 2+1 dimensions. Hitherto, it was not possible to experimentally setup the Haldane model in a crystalline structure \footnote{So far, this model has been only realized in optical lattices \cite{Xu18,Yan18}.}. Instead, another type of QAH insulators was predicted in spin-polarized topological insulators (TIs) like (Hg,Mn)Te quantum wells \cite{Liu08,Qi08} or magnetically doped (Bi,Sb)Te thin films \cite{Yu61,Chang13}. 
Their low-energy physics is captured by the superposition of two Chern insulators \cite{Bernevig06}. Similar to the Haldane model, the gap of one of these Chern insulators can be closed by magnetic doping of the system, whereas at the same time the second Chern insulator remains gapped. In this fine-tuned limit, the gapless Chern insulator realizes the parity anomaly as its contribution to the Hall conductivity is in general non-zero. Hence, the analysis of single Chern insulators, which is the main purpose of the present work, allows to study measurable consequences  of the parity anomaly in a solid state material. In contrast to the Dirac fermions in the Haldane model,  each Chern insulator is characterized by two different parity-breaking mass terms: A conventional Dirac mass, as well as momentum-dependent Newtonian mass.
As the Hall conductivity of a single Chern insulator is unaltered if one takes the parity-symmetric zero-mass limit, both, the Dirac as well as the momentum-dependent mass term, are directly related to the parity anomaly. It was recently shown that the momentum-dependent mass term acts similar to a Wilson fermion in a lattice regularization of a pure Dirac system \cite{Tutschku20}. As such, it ensures an integer- instead of a half-quantized Hall conductivity  associated to a pure Dirac fermion \cite{Lu10,Qi11}.

\noindent
%Since an external magnetic field breaks parity symmetry, the parity anomaly is  a zero magnetic field effect. 
The parity anomaly is a zero magnetic field effect because an external magnetic field breaks the parity symmetry at the classical level. However, it can be shown  that even in quantizing magnetic fields the signatures of the parity anomaly persist. They remain encoded in  the spectral asymmetry \cite{Niemi84}. Below a critical magnetic field, the parity anomaly effectively adds one LL to the entire Hall response of a Chern insulator. Above this field, the magnetic field closes the Dirac mass gap and the system exhibits a conventional QH response \cite{Tutschku19}.

\noindent
All these findings do not incorporate thermal effects. So far, finite temperature signatures in parity anomaly driven systems are restricted to pure Dirac models. 
Calculating the quantum effective action of these systems induces a temperature dependent and thus large gauge non-invariant Chern-Simons term originating from the parity anomaly \cite{Niemi85,dunne97,Dunne98,Sissakian98,Nogueira14,Ma18}.  
While it was shown that this non-invariance is absorbed by 
higher order non-perturbative corrections to the effective action \cite{Deser97,Fosco97,Fosco972,Hott99,Salcedo99,schaposnik99,Beneventano09,Schaposnik2017}, 
this feature still gives rise to a fundamental question:
Does the parity anomaly get renormalized by thermal effects? 
Answering this question is not only relevant for the QAH effect in the materials mentioned above. It is especially important in the case of interfaces between ferromagnetic insulators and three-dimensional topological insulators, where a proximity-induced interface magnetization has been experimentally observed at high temperatures \cite{Moodera-Nature,Moodera-ScienceAdvances-2017}. 
In this case the out-of-plane magnetization causes a gap opening in the interface Dirac spectrum, which induces a parity anomaly 
on the TI surface and a concomitant magnetoelectric torque in the Landau-Lifshitz equation \cite{Nagaosa_PhysRevB.81.241410,Nogueira-Eremin_PhysRevLett.109.237203,Loss_PhysRevLett.108.187201}. 
A similar effect is expected to occur on the surface of the recently discovered antiferromagnetic TI MnBi$_2$Te$_4$ 
\cite{Chulkov-Isaeva,Isaeva_PhysRevX.9.041065}, where the gap in the surface Dirac spectrum is an intrinsic feature of the system.

\noindent
By definition, the parity anomaly only implies the breakdown of parity symmetry at the quantum level. This dictates a certain form of the band-structure, which is  temperature independent \footnote{Rigorously, this statement is only true for small temperatures. For very large temperatures the system can deform, which essentially changes the band-structure. However, this scenario is beyond the scope of our analysis.}.
Hence, the parity anomaly cannot obtain any finite temperature correction. In contrast, the prefactor of the anomaly induced Chern-Simons term in the effective action corresponds to the finite temperature Hall conductivity. 
We calculate this non-dissipative transport coefficient for Chern insulators including both, a Dirac as well as a momentum-dependent mass. We studied these systems in the absence and presence of a magnetic field, as well as with and without particle-hole symmetry. This leads to the following results:
$(i)$ The parity anomaly induces  a topological part in the Hall conductivity which is temperature as well as chemical potential independent and described by the Chern number.
$(ii)$ The non-quantized finite temperature and chemical potential corrections to the Hall conductivity also originate from the parity anomaly, since they also depend on the band-structure. However, they do not depend on its topology, being rather related to the temperature-dependent filling of the valence and conduction bands. %We in particular calculate the finite temperature Hall conductivity for the BHZ model. 
As expected, an increasing Dirac mass counteracts finite temperature effects. 
On the other hand, we show that in the nontrivial phase an increasing Newtonian mass  enhances the finite temperature corrections. 
%$(iii)$ In comparison to a pure Dirac system, the Newtonian mass doubles the topological as well as the thermal corrections to the Hall conductivity. In what follows, we prove these statements analytically.
$(iii)$ In finite magnetic fields,
the thermal LL response renormalizes the parity anomalous part of the Hall conductivity.
In the Dirac mass gap it adds to the otherwise quantized parity anomaly related contribution.\\

\noindent
This work is structured as follows: In 
Sec.~\ref{Sec2}, we discuss the relation of magnetically doped two-dimensional TIs to the parity anomaly and compare these systems to the Haldane model. In this context, we analyze the band-structure of Chern insulators in the absence and presence of an out-of-plane magnetic field and with, as well as without particle-hole symmetry. In Sec.~\ref{Sec3} and Sec.~\ref{Sec4}, we analyze the parity-odd transport of a  Chern insulator for a finite temperature and chemical potential, as well as for zero and finite magnetic fields, respectively. In Sec.~\ref{Sec5} we summarize our results and give an outlook.

\section{Parity Anomaly in a QAH System  Beyond the Haldane Model} \label{Sec2} 
In this work, we consider 2+1 dimensional Chern insulators which are defined by two  different mass terms: A momentum independent Dirac mass $m$, as well as a momentum dependent Newtonian mass term $B \vert \mathbf{k} \vert^2$. The Lagrangian of such an insulator is given by \footnote{Notice, that a similar Lagrangian can be used for the description of 2+1 dimensional superfluid Fermi liquids, as discussed by G.~E.~Volovik in Ref. \cite{volovik1988}.}
\begin{align} \label{lagrangianQAH}
\!  \mathcal{L} & =\bar{\psi} \left(  A \gamma_\mu  k^\mu - m + B k_i k^i  \right) \psi  \ ,
\end{align}
where $\psi$ and $\bar{\psi} \! = \! \psi^\dagger \gamma_0$ are the two-component Dirac spinor and its adjoint,  $\gamma_\mu$=($\sigma_3$,$\im \sigma_2$,$\im \sigma_1)$ are the 2+1 dimensional Dirac matrices, and we consider the metric $g^{\mu \nu}\!=\!\mathrm{diag(+,-,-)}$. 
Here and throughout the manuscript, Greek indices run over the space-time coordinates $\lbrace 0,1,2 \rbrace$, while roman indices run over the spatial components $\lbrace 1,2 \rbrace$ only. 
Moreover, the parameter $A$ is proportional to the Fermi velocity and $\sigma_{1,2,3}$ define the Pauli matrices. Notice, that in comparison to a pure Dirac Lagrangian, the additional Newtonian mass term in 
Eq.~\eqref{lagrangianQAH} breaks the Lorentz symmetry as it only involves spatial momenta.

\noindent
The  first-quantized Hamiltonian associated to Eq.~\eqref{lagrangianQAH} can be derived by a Legendre transformation:
\begin{align} \label{hamchern}
\mathcal{H} = A\left(k_\mathrm{1} \sigma_\mathrm{1} - k_\mathrm{2} \sigma_\mathrm{2} \right)+ \left(m  -  B \alpha \right)\sigma_\mathrm{3} \ .
\end{align}
Here,  we introduced the abbreviation  $\alpha   =   k_\mathrm{1}^2 +  k_\mathrm{2}^2$. Both of the mass terms in Eq.~\eqref{hamchern}, $m$ and  $B \alpha$, break the parity symmetry of the Hamilton. In 2+1 dimensions, parity symmetry is defined as invariance of the theory under \mbox{$\mathcal{P}: \, (x_0,x_1,x_2) \rightarrow (x_0,-x_1,x_2)$}. Consequently, the Dirac as well as the Newtonian mass contribute to the integer Chern number \cite{Lu10,Qi11} \footnote{Adding higher-order momentum dependent mass corrections to the Hamiltonian of a Chern insulator changes its Chern number. To respect the Galilean invariance of the system, any additional mass correction needs to be of even order in momentum. Essentially, the prefactor of the highest order mass correction replaces the Newtonian mass parameter $B$ in the Chern number Eq.~\eqref{CS0}.
Since higher order mass corrections do change the band-curvature, they will also alter the nonquantized Hall response at finite temperatures and chemical potentials. However, they will only change this response quantitatively. In particular, they do not prevent the possible gap closing apart from the $\Gamma$-point which drives the low-energy response studied in this manuscript [cf.~Eq.\eqref{camelback}].}
\begin{align} \label{CS0}
\mathcal{C}_\mathrm{CI}= \left( \mathrm{sgn}(m)+\mathrm{sgn}(B) \right)/2 \ .
\end{align} 
Even in the parity symmetric limit $m,B \rightarrow 0^\pm$, $\mathcal{C}_\mathrm{CI}$ does not vanish for sgn$(m/B)>0$. This effect is known as the parity anomaly of Dirac-like systems in odd space-time dimensions. Initially, the parity anomaly has been predicted for a pure Dirac spectrum in Ref.~\cite{Redlich83}. Due to the absence of a Newtonian mass term, the  Chern number of a pure Dirac system is given by 
$\mathcal{C}_\mathrm{QED}= \pm \mathrm{sgn}(m)/2$ before regularization. Hence, it is half quantized and always non-zero. In contrast, the Chern number in Eq.~\eqref{CS0} is integer quantized and defines two different phases: For $m/B>0$, the system is topologically nontrivial with $\mathcal{C}_\mathrm{CI}=\pm 1$, while for $m/B<0$, the system is topologically trivial with $\mathcal{C}_\mathrm{CI}=0$. In a solid state system with a crystal lattice, Dirac fermions in 2+1 space-time dimensions always  come in pairs \footnote{If the 2+1 dimensional manifold is the boundary theory of a 3+1 dimensional bulk, there could also be an odd number of Dirac fermions living at the boundary. For instance, this is the case at the boundary of 3+1 dimensional topological insulators. However, we are not considering such systems but focus our analysis on pure 2+1 dimensional bulk materials.}. The naive lattice discretization of a pure Dirac fermion leads to a phenomenon called fermion-doubling, which predicts the existence of a second Dirac fermion of opposite Chern number at the edge of the lattice Brillouin zone. Thus, the entire system has Chern number zero and the parity anomaly of a single Dirac fermion cannot be measured. However, in his seminal work \cite{Haldane88}, Haldane
found a way to circumvent this difficulty in a condensed matter system. He proposed a way of how to realize a single Dirac fermion in the bulk spectrum of graphene  by separately manipulating the two Dirac gaps at the $K$ and $K'$ points of the hexagonal lattice structure  via a complex hopping parameter. In particular, this parameter allows to close only one of the Dirac gaps, whereas the other one remains open. Hence, his model suggests a way of how to realize a solid state system which has a single gapless Dirac fermion in 2+1 dimensions but still a non-zero, integer Chern number $\mathcal{C}_\mathrm{HM}=(\mathrm{sgn(m_{\mathrm{K}})-sgn(m_{\mathrm{K}'}))/2}$. 
Here, $m_{\mathrm{K}} $ and $m_{\mathrm{K}'}$ are the Dirac mass terms at the $\mathrm{K}$ and $\mathrm{K}'$ points in graphene, respectively.

\begin{figure}[t]
\flushleft
\includegraphics[scale=.118]{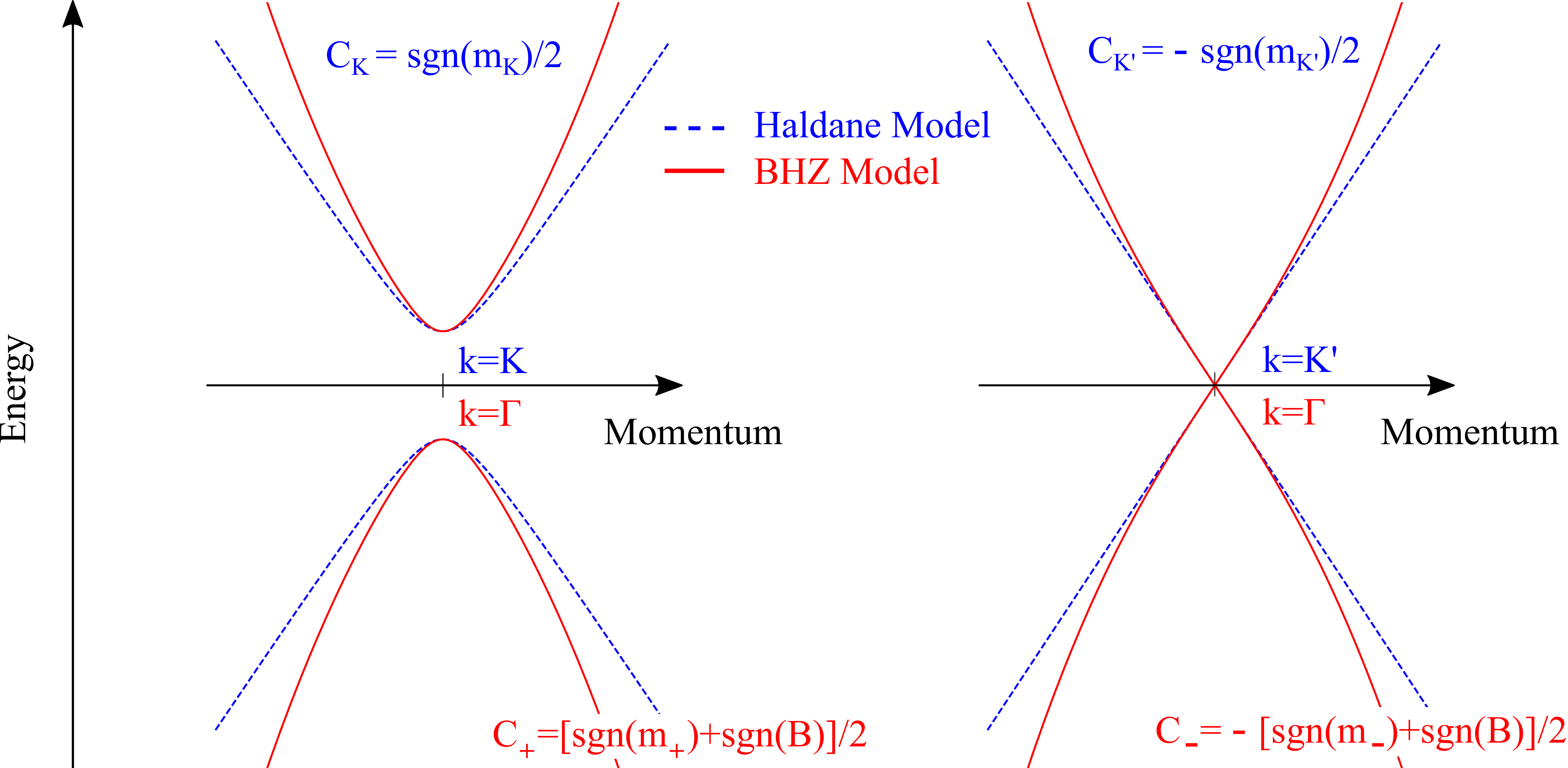}
\caption{Schematic illustration of how a single Dirac fermion or Chern insulator realizes the parity anomaly in the Haldane (blue) or BHZ (red) model. In the Haldane model, both Dirac fermions at the $K$ and $K'$ points of the hexagonal lattice structure contribute $\pm 1/2$ to Chern number, whereas in the QAH phase of the BHZ model only one of the Chern insulators is topologically nontrivial and has a finite Chern number $C=\pm1$. More explanations are given in the text.}
\label{HaldanevsBHZ}
\end{figure}

\noindent
While so far the Haldane model has not yet been realized in a solid state material, a closely related QAH effect has been predicted in two-dimensional 
systems like (Hg,Mn)Te quantum wells or magnetically doped (Bi,Sb)Te thin films.
In the vicinity of the bulk gap, these systems can be effectively described by the Bernevig-Hughes-Zhang (BHZ) model
\begin{align} \label{BHZ}
\mathcal{H}_\mathrm{BHZ}= \begin{pmatrix}
\mathcal{H}_\mathrm{CI}^+ ( \mathbf{k} ) & 0 \\
0 & \mathcal{H}_\mathrm{CI}^{- \ \star} ( \mathbf{-k} ) 
\end{pmatrix}  \ ,
\end{align}
which consists of two copies of the Chern insulators defined in Eq.~\eqref{hamchern}. They are decoupled \footnote{For the scope of this work, bulk inversion asymmetry  terms are unimportant. Therefore, they are neglected throughout the manuscript.} 
and the index $\pm$ defines 
their (pseudo-)spin polarization
\begin{align} \label{hamchernexp}
      \mathcal{H}_\mathrm{CI}^\pm = A\left(k_\mathrm{1} \sigma_\mathrm{1} -  k_\mathrm{2} \sigma_\mathrm{2} \right)  \pm \left(m_\pm   -   B \alpha \right)   \sigma_\mathrm{3}  -D \alpha \sigma_{0}  \ . 
\end{align}
In comparison to Eq.~\eqref{hamchern}, Eq.~\eqref{hamchernexp} also includes a   particle-hole asymmetry $D \alpha \sigma_0$. Since this term is parity-even, it neither contributes to the parity anomaly nor changes the Chern number in Eq.~\eqref{CS0} \cite{JanPaper}. Therefore, let us first consider particle-hole symmetric systems with $D\!=\!0$. The Dirac masses $m_\pm$ of each (pseudo-)spin block in Eq.~\eqref{hamchernexp} can be tuned  by magnetic doping of the system. It is in particular possible to drive one of the Chern insulators in the topologically trivial regime and to close, at the same time, the gap of the second non-trivial Chern insulator. Analogously to the Haldane model, in this scenario the single gapless Chern insulator alone realizes the parity anomaly of a Dirac-like system in 2+1 dimensions. Schematically, this limit is illustrated in Fig.~\ref{HaldanevsBHZ}.

 However, while in the Haldane model both Dirac fermions contribute $\pm 1/2$ to the  Chern number, in the QAH phase of the BHZ model only one of the Chern insulators has Chern number $\mathcal{C}_\mathrm{CI}=\pm 1$. The other one is topologically trivial with $C_\mathrm{CI}=0$. Hence, studying the single Chern insulator in Eq.~\eqref{hamchern} is sufficient to analyze the consequences of the parity anomaly in experimentally realizable systems like 
(Hg,Mn)Te quantum wells or magnetically doped (Bi,Sb)Te thin films.\\

\begin{figure}[t]
\centering
\includegraphics[scale=.2]{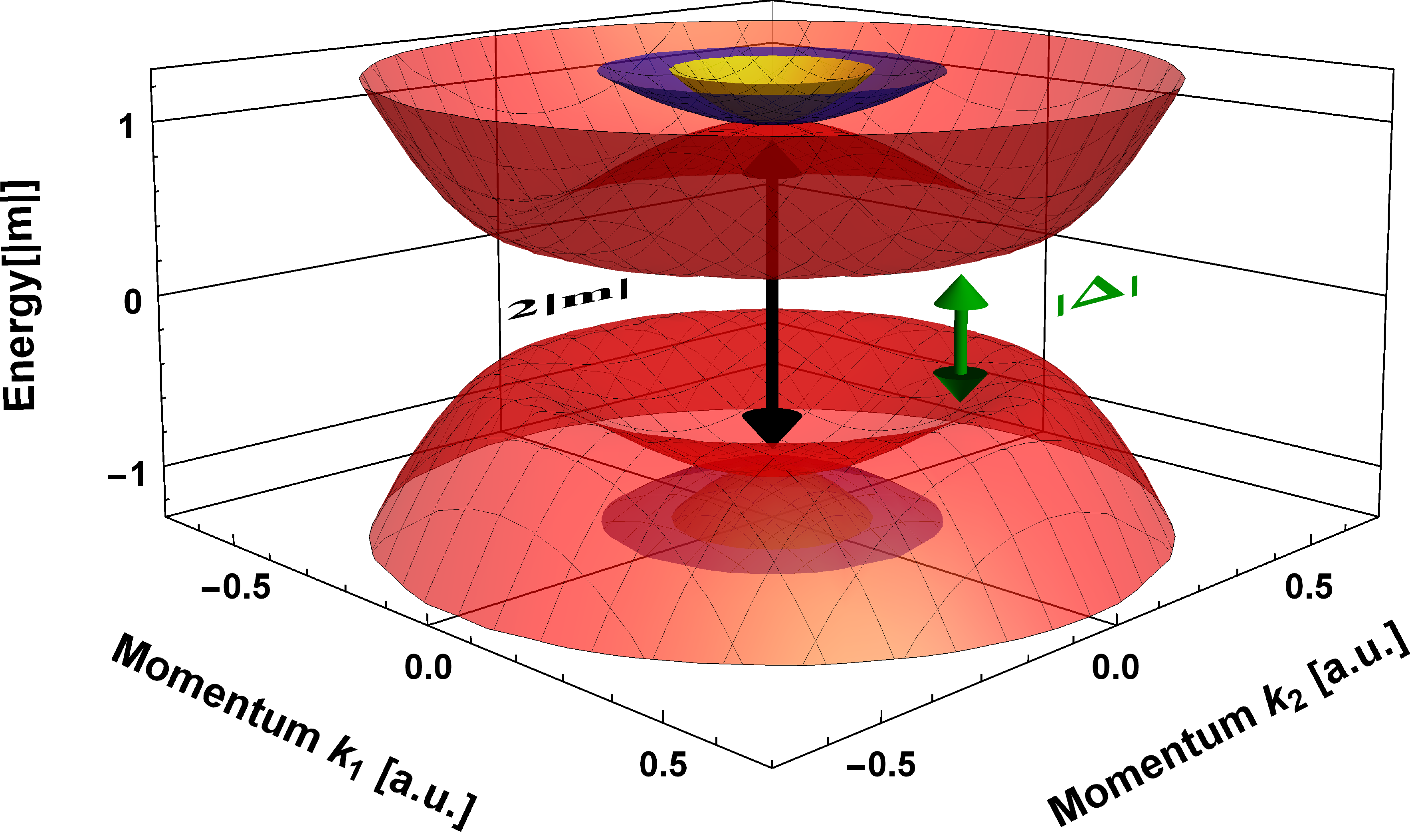}
\vspace{.2cm}
\caption{Band-structure of a Chern insulator for zero magnetic field and $D\!=\!0$. Red and blue curves encode topologically non-trivial phases with $m\!=\!A\!=\!1$, and $B\!=\!5$ (red, camel-back), or $m\!=\!1$, $A\!=\!3$, and $B\!=\!0.1$ (blue).
The yellow curve corresponds to a topologically trivial phase with $m\!=\!1$, $A\!=\!5$, and $B\!=\!-0.1$.  The minimal gap is either defined by $2 \vert m \vert$ at the $\Gamma$-point or by $\vert \Delta \vert$ at $\alpha=\alpha_\mathrm{min}$, indicated by the black or green arrow, respectively.}
\label{fig1}
\end{figure}

\noindent
The spectrum associated to Eq.~\eqref{hamchern} is given by 
\begin{align} \label{bulkspectrum}
\epsilon^\mathrm{\pm}(\alpha)= \pm \sqrt{A^2 \alpha+(m-B \alpha)^2} \ ,
\end{align}
where $\pm$ encodes the conduction and the valence band, respectively. In Fig.~\ref{fig1},  we show the influence of the mass parameters on the  band-structure. Depending on the values for $m$, $B$, and $A$, the  band-structure changes significantly. For $m/B>0$, the system is topologically nontrivial with $\mathcal{C}_\mathrm{CI}=\pm 1$. The minimal gap can be either located at the $\Gamma$-point or at $\alpha_\mathrm{min}\!=\! (2 m B \!-\!A^2)/(2 B^2)$, corresponding to a camel-back structure. Thus, it is  defined by $2 \vert m \vert$ or by the absolute value of
\begin{align} \label{delta}
\Delta=   A \sqrt{4 m B-A^2}/B \ .
\end{align}
Increasing $\vert m \vert$ or $\vert B \vert$  in the nontrivial phase leads to a camel-back structure  if $2mB \! > \! A^2$, associated to $\alpha_\mathrm{min}\!>\!0$. The camel-back gap $\vert \Delta \vert $ increases with $m$ but decreases with $B$. %This property will be crucial when we discuss finite temperature effects.  
For $4 m B \! =\! A^2$, $\Delta$ vanishes and the spectrum simplifies to, 
\begin{align} \label{simplespec}
\epsilon^\mathrm{\pm}(\alpha)=  \pm \vert(m + B \alpha) \vert \ . 
\end{align}
%Thus, the BHZ model has two limits in which one recovers pure Dirac physics. One the one hand, it is the obvious limit $B \rightarrow 0$, on the other hand it is the limit $B \rightarrow 1/(4m)$.\\
For $m/B<0$, the system is topologically trivial. In this case  the minimal gap is  always located at the $\Gamma$-point.\\

\noindent
If we include an out-of-plane magnetic field $H$, a LL spectrum forms if the magnetic  length $l_H\!=\!\sqrt{\hbar/\vert e H\vert}$ is smaller than the system size \cite{Konig08}. For $s\!=\! \mathrm{sgn}(eH)$ and a finite particle-hole asymmetry $D$, one obtains
\begin{align} \label{LLspectrum}
	\epsilon_{n\neq0}^{\pm} & =-s\beta/2 \!-\! n \delta \pm\lambda_n \ , \\ \ \, \epsilon_\mathrm{0}& =s\left( m-\beta/2 \right) \! - \! \delta/2 \ . 
\end{align}
Here, we defined $\alpha \!=\!\sqrt{2}A/l_H$, $\beta\!=\!2B/l_H^2$, $\delta\!=\!2D/l_H^2$,  and  
\begin{align}
\lambda_n= \sqrt{ \alpha^2 n +(m-n\beta - s \delta/2 )^2 } \ \  \mathrm{with} \ \ n \in \mathbb{N}^+ .
\end{align}
As shown in Ref.~\cite{Tutschku19}, $H$ renormalizes the zero-field Chern number $\mathcal{C}_\mathrm{CI}$ in Eq.~\eqref{CS0} to
\begin{align} \label{eq:Intro}
\mathcal{C}_\mathrm{CI}(H)=\left[ \mathrm{sgn} \left(m-B/l_H^2 \right) + \mathrm{sgn} \left( B \right) \right]/2 \ .
\end{align}
%This effect is based on the LL asymmetry and can be derived via  the spectral asymmetry $\eta_\mathrm{H}$ \cite{Niemi85}. 
Hence, a magnetic field counteracts 
the parity anomaly related contribution to the Chern number, and closes the Dirac mass gap at  $H_\mathrm{crit} \! = \! \mathrm{sgn}(eH)m/B$. Beyond this critical magnetic field the parity anomaly vanishes. % In Fig.~\ref{fig1}(b), we plot a LL fan and highlighted the magnetic field $H_\mathrm{crit}$ beyond which the parity anomaly  vanishes \footnote{In quantizing magnetic fields, there is no parity anomaly contribution outside of the Dirac mass gap, since the magnetic field induces a flat LL response.}.\\

\noindent
This highly nontrivial statement deserves further clarification. The parity anomaly of a single Chern insulator is a zero magnetic field effect. It is directly related to the parity breaking elements of the zero-field Hamiltonian in Eq.\eqref{hamchern} and its  band-structure in Fig.~\ref{fig1}.
In quantizing magnetic fields, the  Chern number of each LL only results from the  magnetic field as it only depends on the magnetic length $l_H$. Nevertheless, the parity anomaly still has significant consequences in magnetic fields. Namely, it defines the Chern number in the Dirac mass gap, Eq.~\eqref{eq:Intro}, resulting from the spectral asymmetry of the entire LL spectrum \cite{Tutschku19}. Since for $\vert H \vert > \vert H_\mathrm{crit} \vert$, the spectral asymmetry vanishes, there are no measurable consequences of the parity anomaly beyond this critical value.\\

\noindent 
Above, we have introduced the concept of Chern insulators and have discussed their concrete relation to the parity anomaly in 2+1 space-time dimensions. 
Next, we study finite temperature and density effects on the parity anomaly induced transport by calculating the Hall conductivity in zero, as well as in finite out-of-plane magnetic fields.

\section{Anomaly Induced Transport in Zero Magnetic Field}
\label{Sec3}

In what follows, we calculate the finite temperature Hall conductivity $\sigma_\mathrm{xy}$ corresponding to a particle-hole symmetric Chern insulator at zero magnetic field. This parity-odd and non-dissipative transport coefficient is directly related to the parity anomaly in 2+1 space-time dimensions as it 
does not vanish in the parity symmetric limit $m,B \rightarrow 0^\pm$. In our calculation, we disentangle topological from non-topological contributions to $\sigma_\mathrm{xy}$, the latter originating from thermal effects. For $D\!=\!0$, the finite temperature Hall conductivity  of the Chern insulator in Eq. ~\eqref{hamchern} is given by \footnote{This result is obtained by calculating the vacuum polarization operator or, analogously, the current-current correlation function. In the language of quantum field theory, this corresponds to the evaluation of a one-loop Feynman diagram, whereas in the solid-state community Eq.\eqref{start} results from the Kubo-formalism.} \cite{Tutschku20} 
\begin{align}\label{start}
\sigma_\mathrm{xy}(T,\mu) =  - \dfrac{e^2}{h}   \int \limits_0^\infty   \! \! \mathrm{d}\alpha \,  \dfrac{ A^2(m \! + \! B\alpha)\left[f_\mathrm{v}(T,\mu)  \! - \!    f_\mathrm{c}(T,\mu)\right]}{4 (A^2 \alpha+(m \! - \! B \alpha)^2)^{3/2}}  ,
\end{align}
where $f_\mathrm{c,v}(T,\mu)=[\mathrm{e}^{(\epsilon^\pm(\alpha)-\mu)/ (k_\mathrm{B}T)} +1]^{-1}$ are the conduction and valence  band Fermi functions \cite{Nogueira14}. To disentangle topological from thermal contributions to $\sigma_\mathrm{xy}$, we use that 
\begin{align}
f_\mathrm{v}(T,\mu) =1-  \dfrac{ \Theta(-\epsilon)\, \mathrm{e}^{(\epsilon-\mu)/(k_\mathrm{B}T)}}{\mathrm{e}^{(\epsilon-\mu)/(k_\mathrm{B}T)} +1 }  \ ,
\end{align}
where $\epsilon(\alpha)$ encodes the entire spectrum and $\Theta$ is the Heaviside step function. With this identity,  Eq.~\eqref{start} decomposes into two building blocks,
\begin{align}
\sigma_\mathrm{xy}(T,\mu) & = \sigma^0_{xy} + \sigma_{xy}^1(T,\mu) \ ,
\end{align}
with
\begin{align}
\sigma^0_{xy}  = - \dfrac{e^2}{2 h} \left( \mathrm{sgn}(m)+\mathrm{sgn(B)}\right) \ , \label{parityanomalzyzero}
\end{align}
\vspace{-.3cm}
\begin{align}
\sigma_{xy}^1(T,\mu)   & =  \dfrac{e^2}{h}\int \mathrm{d} \alpha \, \dfrac{A^2(m + B \alpha) \, \mathrm{sgn}(\epsilon)}{4\pi \epsilon^3 } \label{firstcorrzero} 
\, \\ & \times \left(  \dfrac{\Theta(\epsilon)}{\mathrm{e}^{(\epsilon-\mu)/(k_\mathrm{B}T)}   +   1 } + \dfrac{\Theta(-\epsilon)}{\mathrm{e}^{-(\epsilon-\mu)/(k_\mathrm{B}T)}   +   1 } \right) . \nonumber 
\end{align}
Equation~\eqref{parityanomalzyzero} encodes the topological part of the  Hall conductivity. In contrast, Eq.~\eqref{firstcorrzero} defines 
the corrections originating from a finite temperature and chemical potential. These non-topological and thus non-quantized corrections are based on particle-hole excitations of the conduction and valence band. To solve Eq.~\eqref{firstcorrzero}, we use the assumption of a particle-hole symmetric Chern insulator with $D \! = \! 0$. In particular, this implies
\begin{align}
\sigma_{xy}^{1}(T,\mu)  & =  \sigma_{xy}^\mathrm{corr}(T,\mu) + \sigma_{xy}^\mathrm{corr}(T,-\mu) \ .
\end{align}
To determine $\sigma_{xy}^\mathrm{corr}(T,\mu)$ in the energy space, we need to solve Eq.~\eqref{bulkspectrum} for $\alpha$. 
Due to the possible camel-back structure, this leads to two solutions
\begin{align} \nonumber
\alpha_{\pm} & = \alpha_\mathrm{min} \pm  \dfrac{\sqrt{\epsilon^2-\Delta^2}}{ \vert B \vert} \ \ \mathrm{with} \ \ \dfrac{\mathrm{d} \alpha_\pm}{\mathrm{d} \epsilon} = \pm  \dfrac{\epsilon}{ \vert B \vert \sqrt{\epsilon^2- \Delta^2}}  \ .
\end{align}
With these identities, we find the correction
\begin{align} \nonumber
\sigma_{xy}^\mathrm{corr}(T,\mu)  & = \dfrac{e^2}{2h} \Theta[\alpha_\mathrm{min}]    \int \limits_{\vert \Delta \vert}^{\sqrt{m^2}}  \dfrac{A^2+ \dfrac{ 2 \vert B \vert \Delta^2 }{\sqrt{\epsilon^2- \Delta^2}}}{ B  \epsilon^2 \left(\mathrm{e}^{(\epsilon-\mu)/T}   +   1 \right)} \, \mathrm{d \epsilon} \\
 \label{camelback}  & \quad \ + \dfrac{e^2}{4 h}  \int \limits_{\sqrt{m^2}}^\infty \dfrac{A^2+ \dfrac{ 2 \vert B \vert \Delta^2 }{\sqrt{\epsilon^2- \Delta^2}}}{ B  \epsilon^2 \left(\mathrm{e}^{(\epsilon-\mu)/T}   +   1 \right)} \, \mathrm{d \epsilon} \ .
\end{align}
While in Eq.~\eqref{camelback} the second term captures the correction from a monotonic  band-structure, the first term encodes a possible camel-back correction.
For $4mB \!=\!A^2$ with $\Delta\!=\!0$, $\sigma_{xy}^\mathrm{corr}(T,\mu) $ reduces to
\begin{align} \label{QED}
\sigma_{xy}^\mathrm{corr}(T,\mu)   & = \dfrac{m e^2}{h} \int \limits_{\sqrt{m^2}}^\infty \dfrac{A^2}{ \epsilon^2 \left(\mathrm{e}^{(\epsilon-\mu)/(k_\mathrm{B}T)} \! + \! 1 \right)} \ \mathrm{d \epsilon} \ ,
\end{align}
which is twice the QED$_{2+1}$ result with $B = 0$. Analogously to  Eq.~\eqref{parityanomalzyzero}, the Newtonian mass provides a factor of two to the thermal corrections of the QED$_{2+1}$ conductivity. 
For the solution in Eq.~\eqref{QED}, we can define the corrections in terms of the Gamma function $\Gamma(x)$  and the reduced Fermi-Dirac integral $F_j(x,b)$ [App.~\ref{appA}]. In total, this leads to
\begin{align}
\sigma_{xy}^{1}(T,\mu)  & =   \dfrac{e^2}{h} \sum_{s=\pm}  \dfrac{A^2 \, \Gamma(-1)  F_{-2}  \left( \frac{s \mu}{k_\mathrm{B}T}, \frac{\vert m \vert }{k_\mathrm{B}T} \right)}{k_\mathrm{B}T}   \nonumber \\ & \! \! \! \! \! \! \overset{T \ll \vert m \vert}{=} \  \dfrac{e^2}{h} \dfrac{ A^2 \, \Gamma \left(-1,\frac{\vert m \vert}{k_\mathrm{B}T}\right)}{k_\mathrm{B}T} \ . \label{approx}
\end{align}
In Eq.~\eqref{approx}, we approximate the result for low temperatures in comparison to the gap and for zero chemical potential. $\Gamma(s,b)$ is the incomplete Gamma function [App.~\ref{appA}].\\

\begin{figure}[t] 
\centering
\includegraphics[scale=.145]{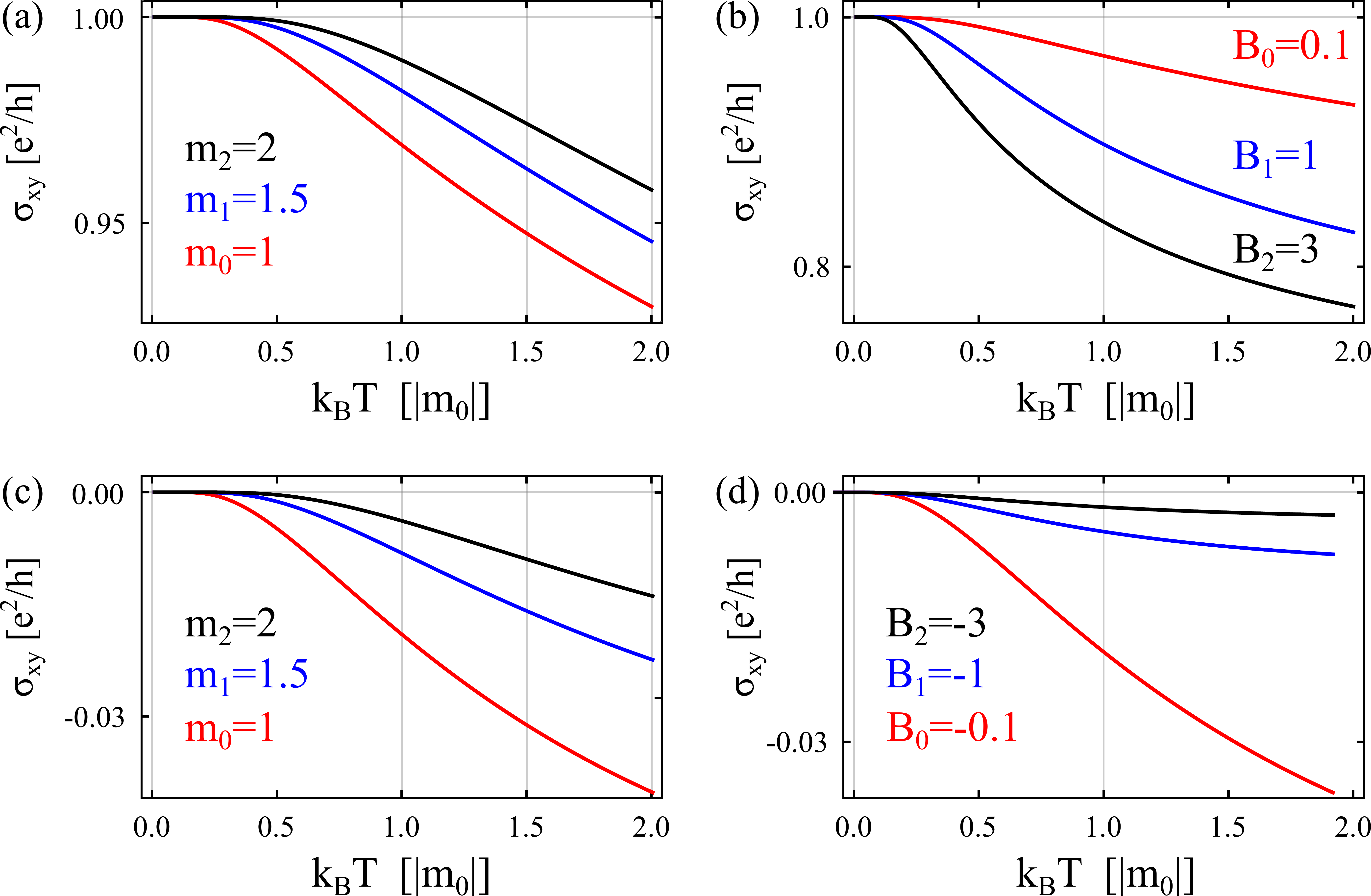}
\caption{Finite temperature Hall conductivity of a Chern insulator with $A\!=\!1$ and $D\!=\!0$. In \textbf{(a)} and \textbf{(c)}, we vary the Dirac mass while $B\!=\! \pm 0.1$, respectively. In \textbf{(b)} and \textbf{(d)}, we vary the Newtonian mass for $m\!=\!1$. 
In all sub-figures we consider zero chemical potential.
Sub-figures \textbf{(a)} and \textbf{(b)} correspond to the the topologically non-trivial regime, while sub-figures \textbf{(c)} and \textbf{(d)} correspond to the topologically trivial regime.}
\label{fig2}
\end{figure}

\noindent
The general correction in Eq.~\eqref{camelback} cannot be expressed via the integral functions above since $\Delta \! \neq \! 0$. In Fig.~\ref{fig2}, we plot the functional dependence of $\sigma_\mathrm{xy}(\mu,T)$ for different choices of $m$ and $B$. While increasing the Dirac mass always counteracts the temperature, increasing $B$ enhances temperature effects in the topologically nontrivial phase. As discussed below Eq.~\eqref{delta}, this originates from the property that $B$ decreases the camel-back gap. Thus, both masses contribute equally to the topological part of the Hall conductivity in Eq.~\eqref{parityanomalzyzero}, while they counteract each other in the thermal corrections, Eq.\eqref{firstcorrzero}, for $m/B>0$.\\
Notice, that even in the topologically trivial phase $m/B<0$ the system has a finite Hall conductivity in the Dirac mass gap [cf.~Figs.~\ref{fig2}(c)~and~(d)].   
This is also directly related to the parity anomaly since it arises from the broken parity symmetry of the  band-structure, which is independent of sgn$(m/B)$. However, in the topologically trivial phase the Newtonian mass cannot generate a camel-back structure. 
In this case both, the Dirac and the Newtonian mass term counteract the finite temperature broadening of the Fermi-Dirac distribution.

\vspace{-.2cm}

\section{Anomaly Induced Transport in Finite Magnetic Field} \label{Sec4}
Having analyzed a particle-hole symmetric Chern insulator at zero magnetic field, we now include a  particle-hole asymmetry and an out-of-plane magnetic field $H$, where the latter gives rise to the LL spectrum in Eq.~\eqref{LLspectrum}.
The Hall conductivity can be calculated by means of Streda's formula via the  expectation value of the charge operator $\langle N \rangle_{T,\mu}$ \cite{Streda82,Niemi85}, yielding
\begin{align} \label{eq41}
\sigma_\mathrm{xy}(T,\mu)=- \dfrac{\partial \langle e N \rangle_{T,\mu}}{\partial H}
= \dfrac{e}{2} \dfrac{\partial  \eta_H}{\partial H} - \dfrac{\partial \langle e N_0 \rangle}{\partial H} \ .
\end{align}
Here,  $\epsilon_\mathrm{z}=-mD/B$ is the  charge neutrality point, and
\begin{align}
\eta_\mathrm{H}  &  = \!  \sum_n \mathrm{sgn}(\epsilon_n \!  - \!   \epsilon_\mathrm{z}) \label{eta}
= - \dfrac{eH}{h} \left[ \mathrm{sgn}(m \!  - \!  \beta/2)   +   \mathrm{sgn}(B)  \right]  , \nonumber
\end{align}
\vspace{-.7cm}
\begin{align}
\langle N_0 \rangle & = 
\! \sum_n \mathrm{sgn}(\epsilon_n \! - \! \epsilon_\mathrm{z}) \! \left[ \dfrac{\Theta(\epsilon_n \!  - \!  \epsilon_\mathrm{z})}{\mathrm{e}^{\frac{\epsilon_n - \mu}{k_\mathrm{B}T}}  \! + \!    1}  \! + \!   \dfrac{\Theta(\epsilon_\mathrm{z}  -   \epsilon_n)}{\mathrm{e}^{-\frac{\epsilon_n - \mu}{k_\mathrm{B}T}} \! + \! 1} \right] . 
\end{align}
The spectral asymmetry $\eta_\mathrm{H}$  counts the difference in the number of conduction and valence band states. Therefore, as long as the band structure is not changed, it is temperature and chemical potential independent and solely carries the information of the topological contribution of the parity anomaly to $\sigma_\mathrm{xy}$ in magnetic fields. This enables the connection between the Hall conductivity and the parity anomaly even at finite magnetic fields. In contrast, $\langle N_0 \rangle$ encodes the thermal LL response, as it defines the thermal occupation of the valence and the conduction band.
Due to the associated  flat dispersion relation,  this response entirely originates from the magnetic field topology  and no more from the parity anomaly. All LLs with $n \! \in \! \mathbb{N}^+$ come in pairs. 
With the degeneracy $\vert e H \vert/h$, their contribution to the charge operator is given by 
\begin{align} \label{higher}
\langle N_0 \rangle_{n \neq 0} = \dfrac{\vert eH \vert }{h} \sum_{n \neq 0,s=\pm}  \dfrac{s}{\mathrm{e}^{\frac{s(\epsilon^s_n - \mu)}{k_\mathrm{B}T}}+1} \ .
\end{align}
Additionally to this conventional LL response for finite $\mu$ and $T$, the zero LL also needs to compensate its contribution to $\eta_\mathrm{H}$ outside of the Dirac mass gap. In particular, it needs to cancel the term $\propto \mathrm{sgn}(m-\beta/2) $ in Eq.~\eqref{eta} for 
$\vert \mu + \delta
/2 \vert \! > \!  \vert m \! - \! \beta/2 \vert $  \cite{Tutschku19}. Since the zero LL can either be part of the conduction or of the valence band, we can  simplify its contribution to $\langle N_0 \rangle$. By using the properties of the hyperbolic tangent, we find for the zero LL with $n=0$
\begin{align*}
\langle N_0 \rangle_{0}   =  - &\dfrac{\vert e H \vert\mathrm{sgn}(\epsilon_0 \!-\! \epsilon_\mathrm{z})}{2 h} \Bigg[ \Theta(\epsilon_0 \!-\! \epsilon_\mathrm{z})  \! \left[ \tanh \! \left( \frac{\epsilon_0 \! - \! \mu}{2k_\mathrm{B}T} \right) \!  - \! 1 \right]  \\ &  \quad \quad \!- \!   \Theta(\epsilon_\mathrm{z} \!-\! \epsilon_0 ) \! \left[ \tanh\left( \frac{\epsilon_0 \! - \! \mu}{2k_\mathrm{B}T} \right) \! + \! 1 \right]  \! \! \Bigg] \ .
\end{align*}
This expression can be simplified further via the identities
\begin{align}
\mathrm{sgn}(\epsilon_0-\epsilon_\mathrm{z}) \! = \! \mathrm{sgn}(\epsilon_0+\delta/2) \!& =\! \mathrm{sgn}(eH) \, \mathrm{sgn}(m\!-\!\beta/2) \nonumber \\
\Theta(\epsilon_0-\epsilon_\mathrm{z})  -\Theta(-\epsilon_0+\epsilon_\mathrm{z}) & =\mathrm{sgn}(\epsilon_0-\epsilon_\mathrm{z}) \nonumber  \\
\Theta(\epsilon_0-\epsilon_\mathrm{z})  +\Theta(-\epsilon_0+\epsilon_\mathrm{z}) & =1 \ .
\end{align}
Eventually, this implies the zero LL contribution
\begin{align} 
\label{zeroll}
\! \langle N_0 \rangle_{0} \!=\! \dfrac{\vert  e H \vert}{2h}  \left[\mathrm{sgn}(eH) \, \mathrm{sgn}(m \! -\! \beta/2)  \! - \!  \tanh \! \left( \frac{\epsilon_0 \! - \!  \mu}{2k_\mathrm{B}T} \right)  \right]   ,
\end{align}
which reduces for $T \rightarrow  0$ and $\bar{\mu}=\mu+D/l_H^2$
to [App.~\ref{appA}]
\begin{align} \label{zeroTlim}
\langle N_0 \rangle_{0}  & =  \nonumber \dfrac{\vert  e H \vert}{2h} \,  \Theta(\vert \bar \mu \vert \! - \!  \vert m \! - \! \beta/2 \vert) \\ & \times \left[ \, \mathrm{sgn}(eH) \, \mathrm{sgn(m \! - \! \beta/2)} + \mathrm{sgn}(\bar \mu)  \, \right]  \ .
\end{align}
For $T \! =\!0$, the zero LL contribution to $\eta_\mathrm{H}$ 
clearly gets compensated outside of the Dirac mass gap.  As expected, finite temperature effects soften this property.\\

\begin{figure}[t] 
\centering
\includegraphics[scale=.16]{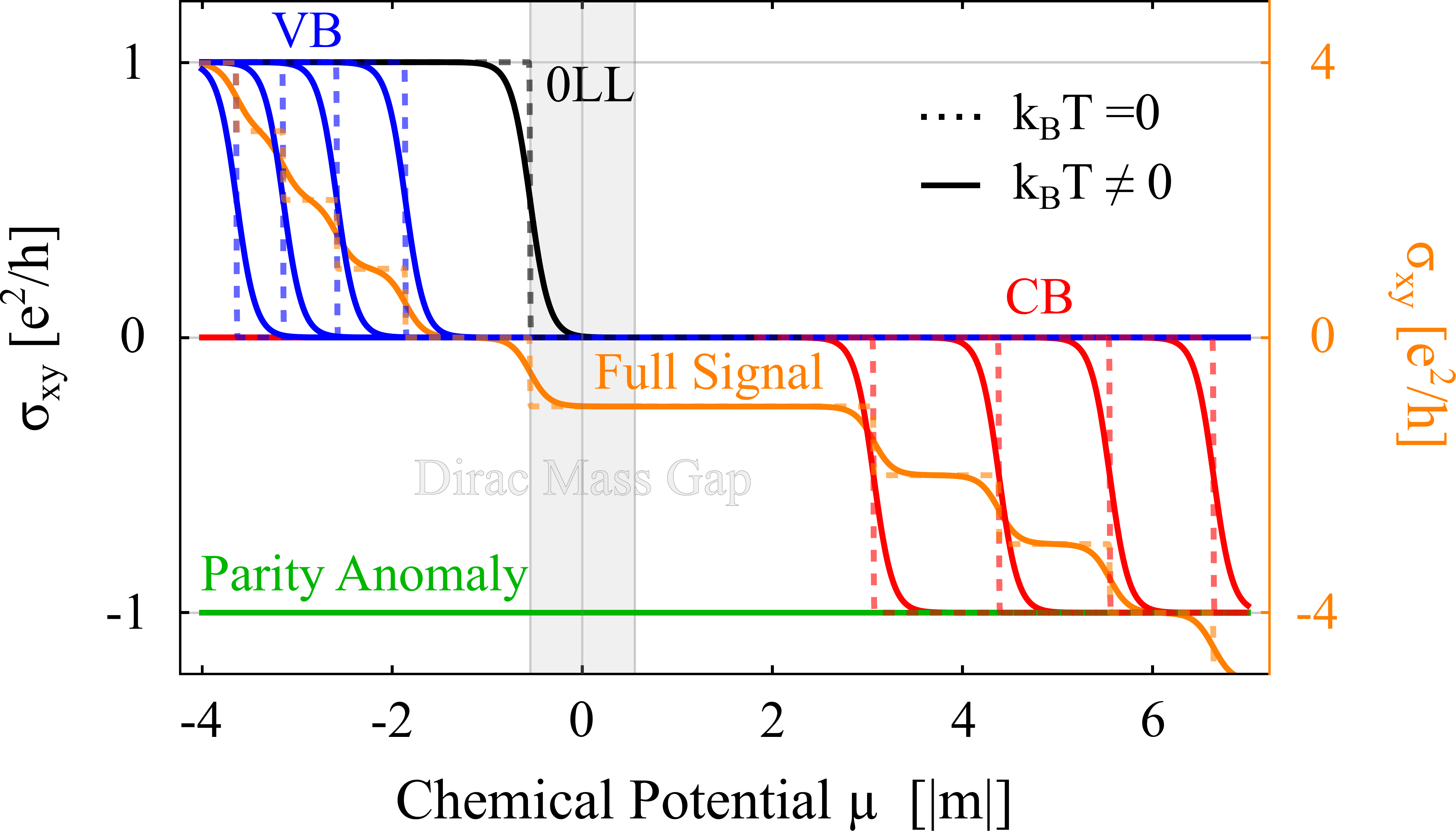}
\caption{Finite temperature Hall conductivity of a Chern insulator 
in a magnetic field $H\!=\!3T$ with:
$A\!=\!1$, $m\!=\!-1$, $B\!=\!-0.1$, and $D\!=\!-0.05$. The response of each valence (blue) and conduction  band (red) LL is shown separately. The zero LL response is illustrated in black, the parity anomaly related contribution is depicted in green. The combined signal is shown in orange. Dashed lines correspond to $k_\mathrm{B}T=0$, solid lines are associated to $k_\mathrm{B}T=0.01$. The Dirac mass gap is shown in grey.}
%\vspace{-.2cm}
\label{fig3}
\end{figure}

\noindent
In Fig.~\ref{fig3}, we used Eq.~\eqref{eq41} to connect the charge operator to $\sigma_\mathrm{xy}$, and plotted Hall conductivity corresponding to the parity anomaly and to each LL, separately. Moreover, we show the entire signal as a function of the chemical potential. While the Hall conductivity contribution related to the parity anomaly is $T$ and $\mu$  independent, each LL comes along with an exponentially suppressed temperature broadening. Consequently, all LLs contribute to the Hall conductivity in the Dirac mass gap. This renormalizes the zero temperature violation of the Onsager relation \footnote{At $T=0$, the only contribution to the Hall conductivity in the Dirac mass gap is given by the parity anomaly in terms of the spectral asymmetry $\eta_H$ [cf.~Eq.~\eqref{eq41}]. Due to Eq.~\eqref{eta} this contribution clearly violates the Onsager relation defined as $\sigma_\mathrm{xy}(-H) = - \sigma_\mathrm{xy} (H)$. In contrast, this identity is fulfilled by any (thermal) LL contribution.} discussed recently in Ref. \cite{Tutschku19}.
%Analogously to the $T\!=\!0$ limit, the $D$-parameter shifts the chemical potential.
Last but not least, let us emphasize that the Hall plateau originating from the parity anomaly is much more robust than  LL plateaus with respect to  finite temperature effects. 
Due to the lack of a zero-LL partner, the parity anomaly response is approximately unaltered \footnote{Due to the exponential suppression of all LL contributions in the Dirac mass gap.} until the chemical potential comes close to the $n\!=\!1$ conduction or valence band LL, depending on the sign of the magnetic field. 
Quantitatively, this means
\begin{align}
\vert \epsilon^\mathrm{+}_{1}-\epsilon_{0} \vert > \vert \epsilon^\mathrm{\pm}_{n+1}-\epsilon^\mathrm{\pm}_{n} \vert \ \quad \  \forall n \in \mathbb{N}^+ \ , 
\end{align}
assuming that the zero-LL is part of the valence band [cf.~Fig.~\ref{fig3}]. Therefore, 
 finite temperature effects firstly smear out the LL steps before they eventually prevent any quantization for the finite temperature 	Hall conductivity.

%\vspace{.2cm}

\section{Summary and Outlook} \label{Sec5}
In this work we analyzed the finite temperature Hall conductivity of two-dimensional Chern insulators under the influence of a chemical potential and an out-of-plane magnetic field.  At zero magnetic field, this quantity originates from the parity anomaly. As such, we were able to show that the parity anomaly is not renormalized by finite temperature effects.	 Instead, it induces two terms of different physical origin in the effective action of a Chern insulator,  which are proportional to its Hall conductivity. The first term is temperature and chemical potential independent, solely 
encoding the intrinsic topological response. The second term specifies the non-topological thermal response of conduction and valence band states.
We showed that in the topologically 
nontrivial phase, an increasing relativistic mass term of a Chern insulator counteracts finite temperature effects, whereas an increasing non-relativistic mass term enhances these corrections. In contrast, both mass terms counteract the finite temperature broadening of the Fermi-Dirac distribution in the topologically trivial phase, as the Newtonian mass cannot cause a camel-back gap in this case. In magnetized II-VI QAH insulators, like (Hg,Mn)Te quantum wells,  these parameters can be tuned by changing the quantum well width, or by changing the concentration of the magnetic dopants. Moreover, we derived the thermal response of a Chern insulator in a magnetic field and clarified its relation to the spectral asymmetry $\eta_\mathrm{H}$. This quantity is a measure of the parity anomaly in  magnetic fields. In particular, we derived in which way the thermal LL response renormalizes the parity anomalous part of the Hall conductivity in magnetic fields. Especially in the Dirac mass gap, this response adds to the otherwise quantized and temperature independent part of the Hall conductivity arising from the parity anomaly. We showed that the  anomalous part of the Hall response in the Dirac mass gap is much more robust than the common LL contributions with respect to finite temperature	effects. Our findings should be experimentally verifiable in QAH insulators such as (Hg,Mn)Te quantum wells, magnetically doped (Bi,Sb)Te thin films, or bilayer structures of three-dimensional topological and ferromagnetic insulators.\\

\noindent
In the future, it would be interesting to extend this analysis to different anomalies in various space-time dimensions. For instance, the chiral, gravitational, and conformal anomaly should not depend on  thermal effects, whereas they necessarily induce a temperature dependence at the effective action level.

\begin{acknowledgments}
We thank J.~Boettcher and R.~Meyer for useful discussions. We acknowledge financial support through the Deutsche Forschungsgemeinschaft (DFG, German Research Foundation), project-id 258499086 - SFB 1170 'ToCoTronics' and SFB 1143 project A5, the ENB Graduate School on Topological Insulators and through the W\"urzburg-Dresden Cluster of Excellence on Complexity and Topology in Quantum Matter - ct.qmat (EXC 2147, project-id 39085490).
\end{acknowledgments}

\bibliographystyle{apsrev4-1}
\bibliography{newBib}

\appendix

\section{Appendix} \label{appA}

\noindent
The reduced Fermi-Dirac integral is defined by
\begin{align}
F_j(x,b)= \dfrac{1}{\Gamma(j+1) }\int \limits_{b}^\infty \mathrm{d} t \, \dfrac{t^{j} }{    \mathrm{e}^{t-x} +1  }  \ .
\end{align}
The incomplete $\Gamma$-function is defined by
\begin{align}
\Gamma(s,b)= \int_b^\infty \mathrm{d}t \, t^{s-1} e^{-t} \ .
\end{align}
Let us comment on how to derive Eq.~\eqref{zeroTlim} from  Eq.~\eqref{zeroll}. In the zero temperature limit, the hyperbolic tangent in Eq.~\eqref{zeroTlim} becomes a sign-function, $\lim_{T \rightarrow 0}\mathrm{tanh}(x/T)=\mathrm{sgn}(x)$. Due to this property, we need to distinguish two cases. The chemical potential is either located inside $(i)$ or outside $(ii)$ of the Dirac mass gap:
\begin{align}
(i) \quad \vert \mu + \delta/2 \vert    &< \vert m-\beta/2 \vert  \\
(ii) \quad \vert \mu + \delta/2 \vert    &> \vert m-\beta/2 \vert \  .  
\end{align}
For case $(i)$, the hyperbolic tangent in Eq.~\eqref{zeroTlim} reduces to $\mathrm{sgn}(eH)\mathrm{sgn}(m \! -\! \beta/2)$ and consequently leads to $\langle N_0 \rangle_{0}=0$. Instead, for case $(ii)$, it reduces to $\mathrm{sgn(\mu + \delta/2)}$, eventually implying Eq.~\eqref{zeroTlim}. While the first, temperature independent term in  Eq.~\eqref{zeroTlim} describes the asymmetry of the zero LL with respect to zero energy, the second term encodes its temperature-dependent response. This term ensures that at $T \! = \!0$ the zero LL only contributes outside of the Dirac mass gap, exactly compensating its contribution to the spectral asymmetry. As expected, this property becomes softened by finite temperature effects.

\end{document}